\documentclass[fleqn,usenatbib]{mnras}

\usepackage{times}
\usepackage{url}
\usepackage{epsfig}
\usepackage[flushleft]{threeparttable}
\usepackage{amssymb}
\usepackage{amsmath}
\usepackage{pdfpages}
\usepackage{todonotes}
\usepackage{array}
\usepackage{comment}
\usepackage{pdflscape}
\usepackage{siunitx}

\usepackage[T1]{fontenc}

\newcommand{\refbf}{} 

\title[SMSS-BLAP-1]{Discovery of A New Blue Large-Amplitude Pulsator in the SkyMapper DR2: SMSS J184506.82-300804.7}

\author[Chang et al.]{Seo-Won Chang$^{1,2,3}$, Christian Wolf$^{4,5}$, Christopher A. Onken$^{4}$, Michael S. Bessell$^{4}$ \\
$^{1}$SNU Astronomy Research Center, Seoul National University (SNU), 1 Gwanak-rho, Gwanak-gu, Seoul 08826, Korea\\
$^{2}$Astronomy Program, Dept. of Physics \& Astronomy, SNU, 1 Gwanak-rho, Gwanak-gu, Seoul 08826, Korea\\
$^{3}$Center for the Gravitational-Wave Universe, SNU, 1 Gwanak-rho, Gwanak-gu, Seoul 08826, Korea \\
$^4$Research School of Astronomy and Astrophysics, Australian National University, Weston Creek ACT 2611, Australia \\
$^5$Centre for Gravitational Astrophysics 
(CGA), Australian National University, Acton ACT 2601, Australia
\\
}

\begin{document}
\label{firstpage}
\pagerange{\pageref{firstpage}--\pageref{lastpage}}
\date{draft \today}

\maketitle

\begin{abstract}
We report the discovery of a new Blue Large-Amplitude Pulsator, SMSS J184506-300804 (SMSS-BLAP-1) in Data Release 2 of the SkyMapper Southern Sky Survey. We conduct high-cadence photometric observations in the $u$ band to confirm a periodic modulation of the lightcurve. SMSS-BLAP-1 has a $\sim$19-min pulsation period with an amplitude of 0.2 mag in $u$ band, and is similar to the classical BLAPs found by OGLE. From spectroscopic observations with the Wide-Field Spectrograph on the ANU 2.3m telescope, we confirm it as a low-gravity BLAP: best-fit parameters from the non-LTE Tlusty model are estimated as {\refbf $T_\mathrm{eff}$ = 29,020$^{+193}_{-34}$ K, $\log g$ = 4.661$^{+0.008}_{-0.143}$ (cm s$^{-2}$), and $\log$ n(He)/n(H) = -2.722$^{+0.057}_{-0.074}$ dex}. However, our BLAP exhibits a very He-deficient atmosphere compared to both low- and high-gravity BLAPs, {\refbf which have $\log$ n(He)/n(H) in the range -0.41$\sim$-2.4.}
\end{abstract}

\begin{keywords}
stars: oscillations --- stars: variable: general ---  stars: fundamental parameters – stars: individual: SMSS J184506.82-300804.7

\end{keywords}


\section{Introduction}\label{sec:intro}
Blue Large-Amplitude Pulsators (BLAPs) are a rare and striking class of short periodic (from a few to 40 minutes), luminous hot subdwarf stars with exceptionally large-amplitude flux variations of 20--40\% and unclear origin. The saw-tooth shape of their phased light curve is not atypical compared to other classical pulsators (e.g., $\delta$ Scuti stars or Cepheids) and their mono-periodic nature suggests an iron-group opacity origin \citep{Pietrukowicz2017NatAs...1E.166P, McWhirter2020MNRAS.496.1105M}. These subdwarf pulsators are significantly blue in colour because of high effective temperatures of $T_\mathrm{eff}$ = 26,000--34,000 K (e.g., \citealt{Pietrukowicz2017NatAs...1E.166P, Kupfer2019ApJ...878L..35K,Pigulski2022A&A...663A..62P,Lin2022NatAs.tmp..217L}). The spectrum of BLAPs also exhibits hydrogen and helium absorption lines corresponding to high effective temperatures. Spectra suggest surface gravity levels of $\log g$ (cm s$^{-2}) \approx 4\ldots 5$, but there is an additional class of BLAPs with surface gravity over the range of 5 $< \log g <$ 6 \citep{Kupfer2019ApJ...878L..35K}. 

The formation channel and evolutionary path of BLAPs remains open. As stars evolve through the instability strip, the rate of period change $\dot{P}/P$ should be associated with gradual changes in temperature and radius, which gives an important hint for potential evolutionary pathways. \citet{Pietrukowicz2017NatAs...1E.166P} have shown that the pulsation period of eleven classical BLAPs is changing very slowly with $\dot{P}/P=-2.85\times 10^{-7} \mathrm{yr}^{-1}$ to $+7.65\times 10^{-7} \mathrm{yr}^{-1}$ (mostly positive) over about 1.5 decades. Two models for BLAPS have been proposed: core helium burning and hydrogen-shell burning. The core helium burning model requires significant mass loss and an inflated envelope to explain the observed pulsations, which are powered by helium core stars with a mass of $\sim 1.0 M_{\odot}$. The hydrogen-shell burning model can also explain the properties of BLAPs, but with less severe mass loss and helium core stars with a mass of $\sim 0.3 M_{\odot}$. Additionally, the period changes of another three BLAPs were traced with further OGLE IV data, yielding values down to $\dot{P}/P = -19.23\times 10^{-7} \mathrm{yr}^{-1}$ \citep{Wu2018MNRAS.478.3871W}. In their new model, BLAPs are more likely to be core helium burning stars with masses $\sim$ 0.7--1.0$M_{\odot}$. Interestingly, \citet{Lin2022NatAs.tmp..217L} reported a firmly confirmed BLAP with a rate of period change of $\dot{P}/P$ = 2.2$\times$10$^{-6}$ yr$^{-1}$, which is an order of magnitude larger than previous ones and has a positive value. Such evolutionary time-scales are not compatible with either of the two leading models -- low-mass, helium-core pre-white dwarfs (e.g., \citealt{Romero2018MNRAS.477L..30R,Kupfer2019ApJ...878L..35K,Byrne2020MNRAS.492..232B}) or more massive, core helium-burning stars with very thin hydrogen envelope. It brings up shell helium-burning subdwarfs (SHeB) as an additional solution to explain this unusual rate of period change. \citet{Xiong2022A&A...668A.112X} push this SHeB model a step further and predict that the positive rate of period change may evolve to negative.

Different model populations that likely behave as BLAPs prefer to reside in binary systems, even with a diversity in formation channels. With no effect of binary interaction (e.g., mass transfer or common envelope ejection), these BLAP-like stars would take longer than the Hubble timescale for reaching out to characteristic regions that are occupied by most known BLAPs. Until recently, however, a companion star was identified solely in the one brightest BLAP (\(V\) $\approx$ 11) with the measurement of an orbital period of $P_\mathrm{orb}\sim$ 23 days \citep{Pigulski2022A&A...663A..62P}. The presence of the companion was further confirmed by the overall shape of their composite spectral energy distribution (SED), revealing it as a late B-type primary component in the main-sequence (MS) phase. An outstanding question for remaining BLAPs that have no signs of binarity is that where all the companions have gone. In a recent test of binary evolution pathways for BLAPs, the bulk of the models with surviving companions have quite faint objects as either low-mass main-sequence stars or white dwarfs \citep{Byrne2021MNRAS.507..621B}. When these companions are in the BLAP phase of binary evolution, the orbital periods show a multimodal distribution peaked around 1.2~d and 40~d where the latter is a dominant one. The extra long-term monitoring effort is thus unavoidable because it is the only promising way to reveal their existence. The only known case is the BLAP discovered by \citet{Lin2022NatAs.tmp..217L}, which has a long orbital period of $\sim$1576 days and is thought to have formed through Roche lobe overflow. Unsurprisingly, there is also a route to form single BLAPs as survivors of the supernova explosion of their primary stars \citep{Meng2020ApJ...903..100M,Meng2021MNRAS.507.4603M}. This channel begins with the Type Ia supernova explosion of a carbon-oxygen white dwarf (CO WD) that impacts its red giant companion. The companion loses its envelope and evolves into a BLAP. This channel also produces BLAPs with a mass range of 0.65--1.0$M_{\odot}$, similar to the model of \citet{Wu2018MNRAS.478.3871W}. Very recently, the formation of BLAPs through mergers (HeWD+MS) has been proposed as an alternative explanation for the absence of binary companions \citep{Zhang2023arXiv231107812Z}. 

We are still at an early stage of understanding the reality of the BLAPs because more samples with long-term photometry and/or precise spectroscopic measurements are needed for defining the full parameter space occupied by the whole population. Observationally, there are difficulties in searching for BLAPs in our Galaxy: (i) Nearly all confirmed BLAPs are located in low Galactic latitude regions ($|b| < 10^{\circ}$; mean $E(B-V)$ = 0.97$\pm$0.81) where their observed photometry deviates from their intrinsic colour due to interstellar extinction and source confusion. (ii) It is not easy to obtain an unbiased estimate of such short pulsational periods without support from minute-cadence (even sub-minute cadence) observations. These difficulties may lead to a large discrepancy between model predictions and observations so far. Model predictions suggest that BLAP-like objects should be common: up to 12,000 low-mass helium-core pre-white dwarfs are expected to be visible as BLAPs through either common envelope evolution or Roche lobe overflow channels \citep{Byrne2021MNRAS.507..621B} or 3,500$\sim$70,000 SHeB sdB stars can be formed through both evolutionary channels \citep{Xiong2022A&A...668A.112X}. Moreover, \citet{Meng2020ApJ...903..100M} estimated a rough number density of 750--7500 BLAPs as surviving companions of SNe Ia. But in reality only seven stars have been further confirmed as BLAPs after the first discovery of 14 BLAPs by the Optical Gravitational Lensing Experiment survey \citep{Pietrukowicz2017NatAs...1E.166P,Kupfer2019ApJ...878L..35K, McWhirter2022MNRAS.511.4971M, Pigulski2022A&A...663A..62P,Lin2022NatAs.tmp..217L}. At the time of submission, another 43 BLAPs were detected by the OGLE-IV survey \citep{Borowicz2023AcA....73....1B}. 

In this work, we present a new member of the class of BLAPs and discuss its photometric and spectroscopic properties that clarify its nature as a low-gravity object. In Section 2 we introduce our newly discovered object and provide additional details on the subsequent photometric and spectroscopic follow-up observations. In Section 3 we measure the pulsation period from light curves and stellar parameters from spectra. In Section 4 we compare our new object with the known population, especially in terms of location in the Galaxy and kinematics. We close with a summary in Section 5.


\begin{figure}
\includegraphics[width=\linewidth]{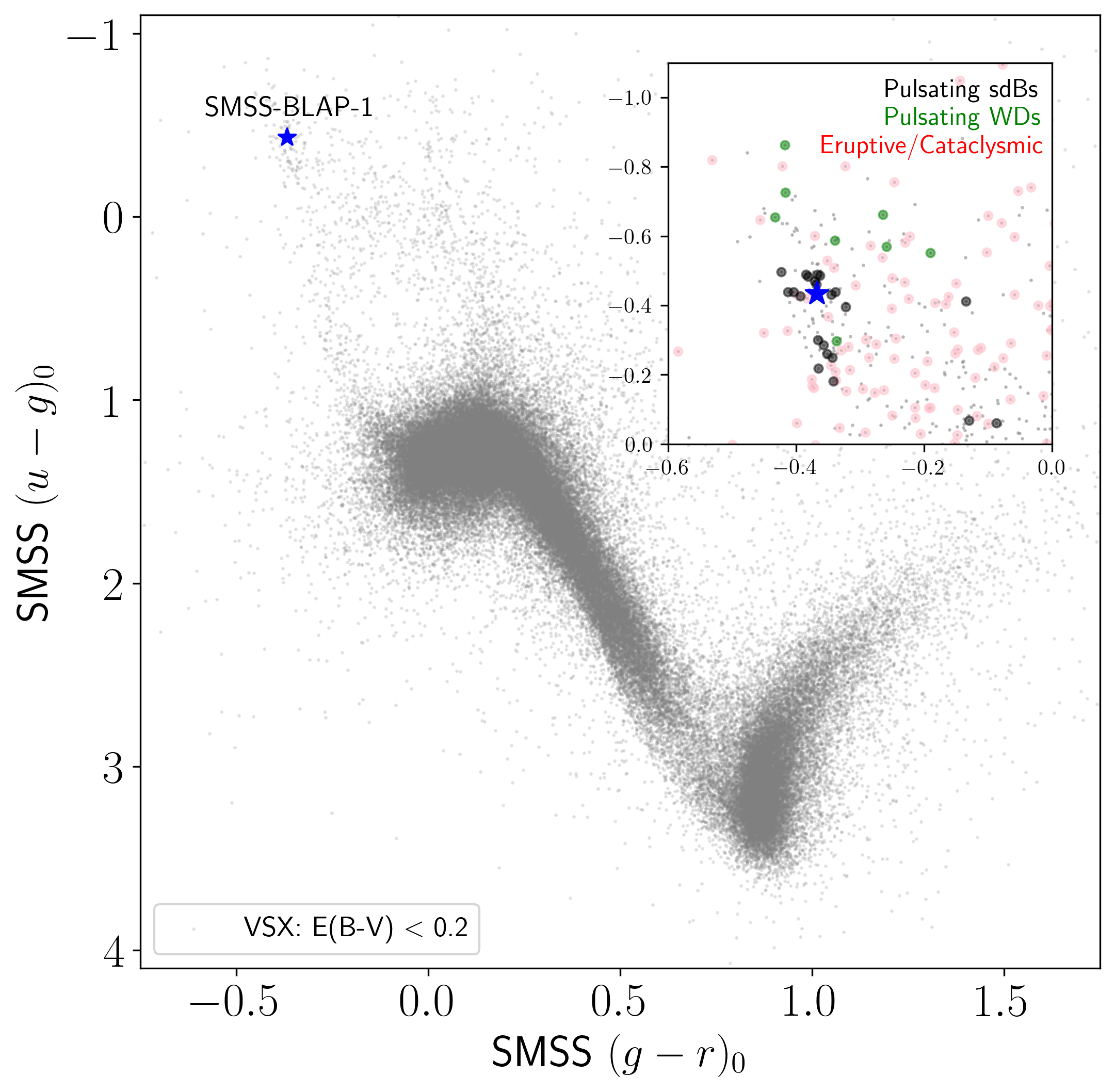}
\includegraphics[width=\linewidth]{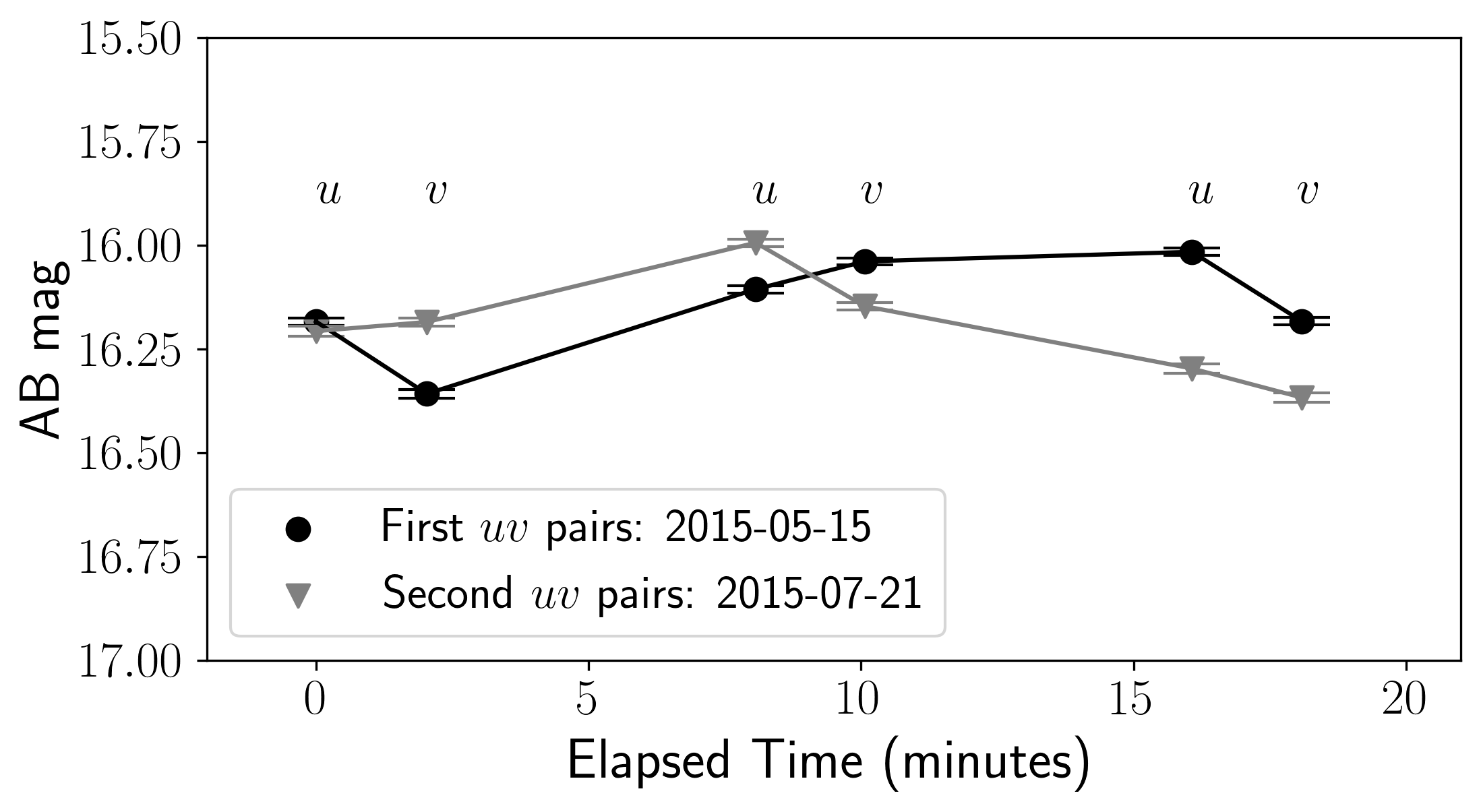}
\caption{Top: Extinction-corrected colour-colour diagram of SMSS-BLAP-1 (blue star) and 127,660 known variable stars (grey points). Bottom: Sparsely-sampled $uv$ light curves of SMSS-BLAP-1 observed over a 20-minute period in two different epochs.}
\label{fig:FactSheet of SMSS-BLAP-1}
\end{figure}

\section{OBSERVATIONS}\label{sec:observation}
\subsection{Discovery} 
SMSS J184506.82-300804.7 (SMSS-BLAP-1)\footnote{\url{https://skymapper.anu.edu.au/object-viewer/dr2/225223741/}} is a relatively bright (\(G_\mathrm{DR3}\)=16.476), very blue object with colour index \(G_\mathrm{BP}\)-\(G_\mathrm{RP}\) = -0.248 in the constellation of Sagittarius ($l, b$ = 5.035, -11.987). The most recent parallax-based distance using Gaia DR3 data is approximately 7063$^{+2058}_{-1622}$ pc \citep{GAIA2022arXiv220800211G,Bailer-Jones2021AJ....161..147B}. While dust is present along this line of sight, the upper limit for the total extinction from the 2D dust emission maps of \citet{Schlegel1998ApJ...500..525S} is modest with $E(B-V) = 0.144$. This object was originally classified as a hot subdwarf candidate based on the Gaia Data Release 2 (Gaia DR2) catalogue and multi-band photometric catalogues \citep{Geier2019A&A...621A..38G}, but there was no further verification of its nature. {\refbf Additionally, it is worth noting that this object is not present in the recently updated DR3 catalogue by \citet{Culpan2022A&A...662A..40C}.}

We first identified SMSS-BLAP-1 during a search for short-term variable blue stars in the Second Data Release of the SkyMapper Southern Survey (SMSS DR2, \citealt{Onken2019PASA...36...33O}). DR2 includes the first release of images from the Main Survey component, which is designed to visit a patch of sky three times within 20 minutes in the \(uv\)-band pairs with an exposure time of 100 seconds per image. The central wavelengths and width of these two filters are 350/43~nm and 384/31~nm, respectively. These multi-epoch time series observations can be useful for finding BLAP-like candidates in the Southern sky that are not hidden behind dust. In our initial search we found four candidates with variability on the timescale of the Main-Survey colour sequence: one of these is SMSS-BLAP1, one turned out to be a Roche-lobe filling sdO+white dwarf binary with a 3.5 hour period \citep{Li2022MNRAS.515.3370L}, and two objects require further observations to settle their nature. 

Figure \ref{fig:FactSheet of SMSS-BLAP-1} shows the location of SMSS-BLAP-1 and 0.1 million known variable stars (VSX: \citealt{2006SASS...25...47W,2017yCat....102027W}) in the SMSS \((u-g)_{0}\) vs. \((g-r)_{0}\) colour-colour diagram corrected for extinction\footnote{\url{https://skymapper.anu.edu.au/filter-transformations/}}. Most of un-reddened variable stars are confined to a tight locus in colour-colour space, while the location for our BLAP is far away from the blue edge in \((u-g)_{0}\). As shown in the insert figure, variable objects in the colour–colour plane around SMSS-BLAP-1 (-0.368, -0.435) are occupied by pulsating hot subdwarf stars (V361 Hya and V1093 Her types) and pulsating white dwarfs (ZZ Cet types). There is also non-negligible population showing sudden eruptions or outbursts (GCAS or UG types), which can mimic a BLAP-like colour by a random chance in sparsely-sampled DR2 light curves. Even in DR2, we see relatively large changes ($\sim$0.22--0.3 mag) in the overall \(uv\) brightness over less than 20 minutes (see bottom panel of Figure \ref{fig:FactSheet of SMSS-BLAP-1}). Interestingly, this object was included in a list of promising Large-Amplitude Variables (LAVs) and classified as Gaia DR2 LAV candidate with amplitude larger than $\sim$0.2 mag in the \(G\) band \citep{Mowlavi2021A&A...648A..44M}. 

\subsection{SkyMapper High-Cadence Photometry}
We conducted a high-cadence monitoring of the BLAP candidates to verify their pulsation periods. We adopted a repeated filter sequence, alternating between a 100-sec $u$ band and a 40-sec $i$ band exposure. Altogether, we obtained 283 $u$ band and 240 $i$ band images with the 1.35-m Australian National University (ANU) SkyMapper Telescope at Siding Spring Observatory (SSO), from 10 July 2019 to 27 August 2019 (15 nights) under a Discretionary Time program. Most images were obtained during bad-seeing time with an airmass range of 1--1.4, having a mean seeing of 4.36$\pm$0.45 arcseconds. Individual photometric measurements are extracted using the SkyMapper Science Data Pipeline \citep{Wolf2018PASA...35...10W,Onken2019PASA...36...33O}. To construct the final magnitude estimates, we use a 1D-PSF photometry output based on a sequence of aperture magnitudes as produced for the data releases of the SkyMapper Southern Survey.

\begin{figure}
\includegraphics[width=\linewidth]{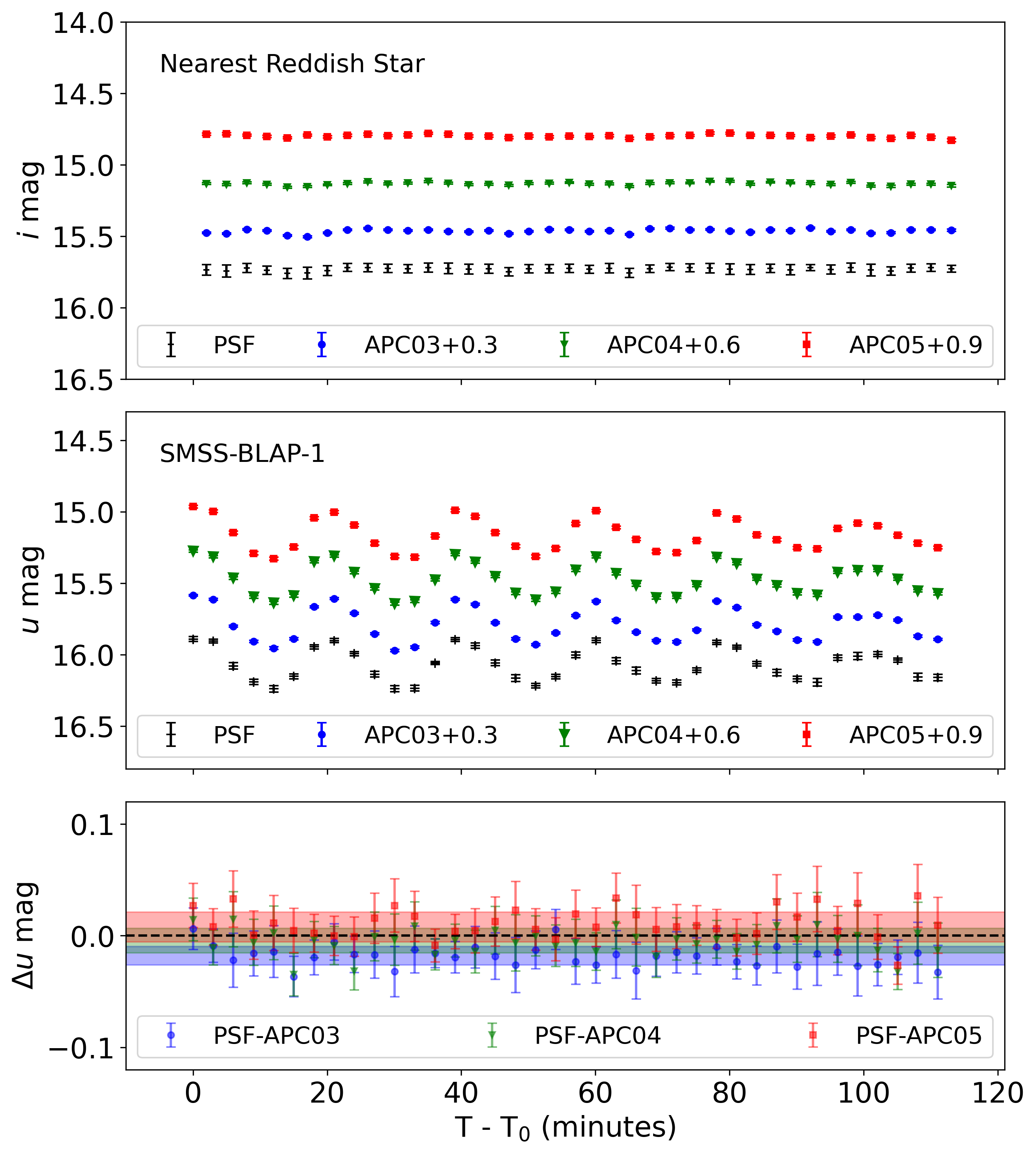}
\caption{Example high-cadence photometric observations of SMSS-BLAP-1 ($u$-filter) and its nearest reddish star ($i$-filter) obtained from the SkyMapper telescope on 10 July 2019. To check the effect of the neighbour star, we compare the magnitudes measured in 3 (blue), 4 (green), and 5 (red) arcsecond apertures with a curve-of-growth correction. For clarify purposes, we shift the light curves vertically relative to their baseline light curves measured by the weighted median PSF magnitude (black). In any photometry aperture, SMSS-BLAP-1 exhibits a clear short-term variability compared to as shown in Fig \ref{fig:FactSheet of SMSS-BLAP-1}, while the nearest star turns out to be non-variable.}
\label{fig:SMSS High-Cadence photometry}
\end{figure}

One issue in determining photometric variability periods is that the 1D-PSF photometry of the candidate can be affected by nearby objects during poor seeing conditions. The nearest DR2 object visible on SMSS images is a star at 3.6\arcsec distance to the South-West. This nearby star is at least 0.5 mag brighter than our candidate in Gaia $G$-Band but has a red colour (\(G_\mathrm{BP}\)-\(G_\mathrm{RP}\) = 0.987), which makes it more impacting at longer wavelengths. In Figure \ref{fig:SMSS High-Cadence photometry}, we compare several magnitudes measured in apertures with diameters of 3, 4, and 5 arcseconds that are corrected for the growth curve of the expected PSF at the object location. For both stars, all these apertures exhibit nearly identical magnitudes. The RMS scatter in the both lightcurves is negligible, with a value around 0.01 mag.

\subsection{ANU 2.3m Spectroscopy}
Spectroscopic observations of the candidate were carried out with the Wide Field Spectrograph (WiFeS; \citealt{Dopita2007Ap&SS.310..255D,Dopita2010Ap&SS.327..245D}) on the ANU 2.3m-telescope at Siding Spring Observatory. WiFeS is an image slicer-based integral-field spectrograph that records optical spectra over a contiguous field of view of 38$\arcsec$$\times$25$\arcsec$. This spatial field is divided into twenty-five 1$\arcsec$ wide slitlets with 0.5$\arcsec$ sampling along the 38$\arcsec$ lengths. As it is a dual-channel device, each camera for the blue and red channels is equipped with 4096$\times$4096 pixel CCD detector. We use the two high resolution gratings (B7000 and R7000) to cover the wavelength range 3,290--7,060\SI{}{\angstrom} with a spectral resolution of R=7,000. 

We obtain nine 600-sec spectra with WiFeS on 9 June 2019 between 19$^{h}$12$^{m}$ and 20$^{h}$46$^{m}$ sidereal time. The blue and red spectra cover the wavelength range of 3300--4370\SI{}{\angstrom} and 5300--7040\SI{}{\angstrom}, respectively. The sliced 2D spectra were processed into wavelength-calibrated 1D spectra using FIGARO \citep{Shortridge1993ASPC...52..219S}; FIGARO techniques were employed for bias subtraction, cosmic-ray removal, flat fielding, slice extraction, field distortion correction, and sky subtraction. The 1D spectra were corrected for extinction and flux calibrated using the spectrophotometric standard star GJ754.1A (DBQA5; \citealt{Bohlin2014PASP..126..711B,Bohlin2022stis.rept....7B}).


\section{Analysis}\label{sec:analysis}
\begin{figure}
\includegraphics[width=\linewidth]{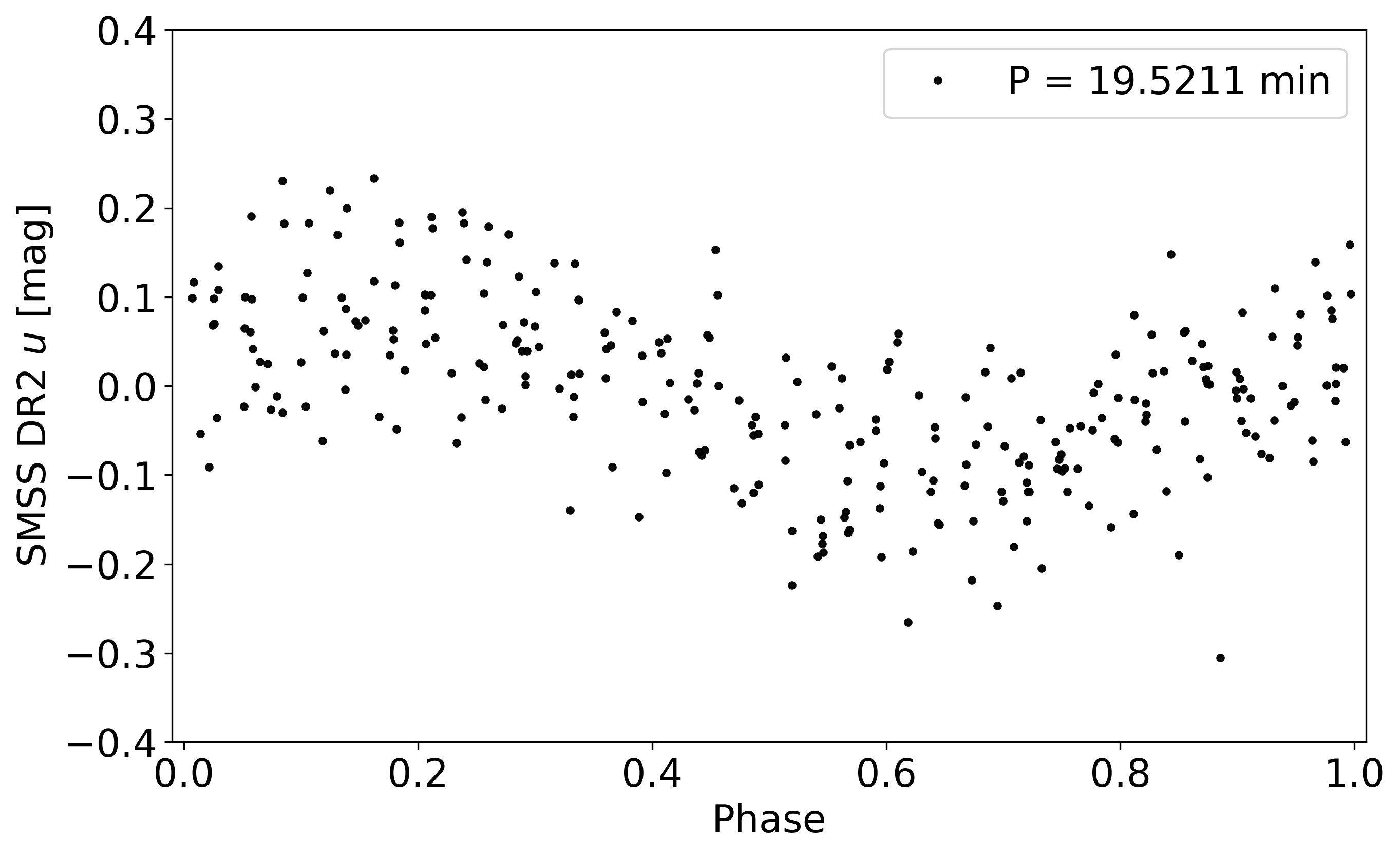}
\includegraphics[width=\linewidth]{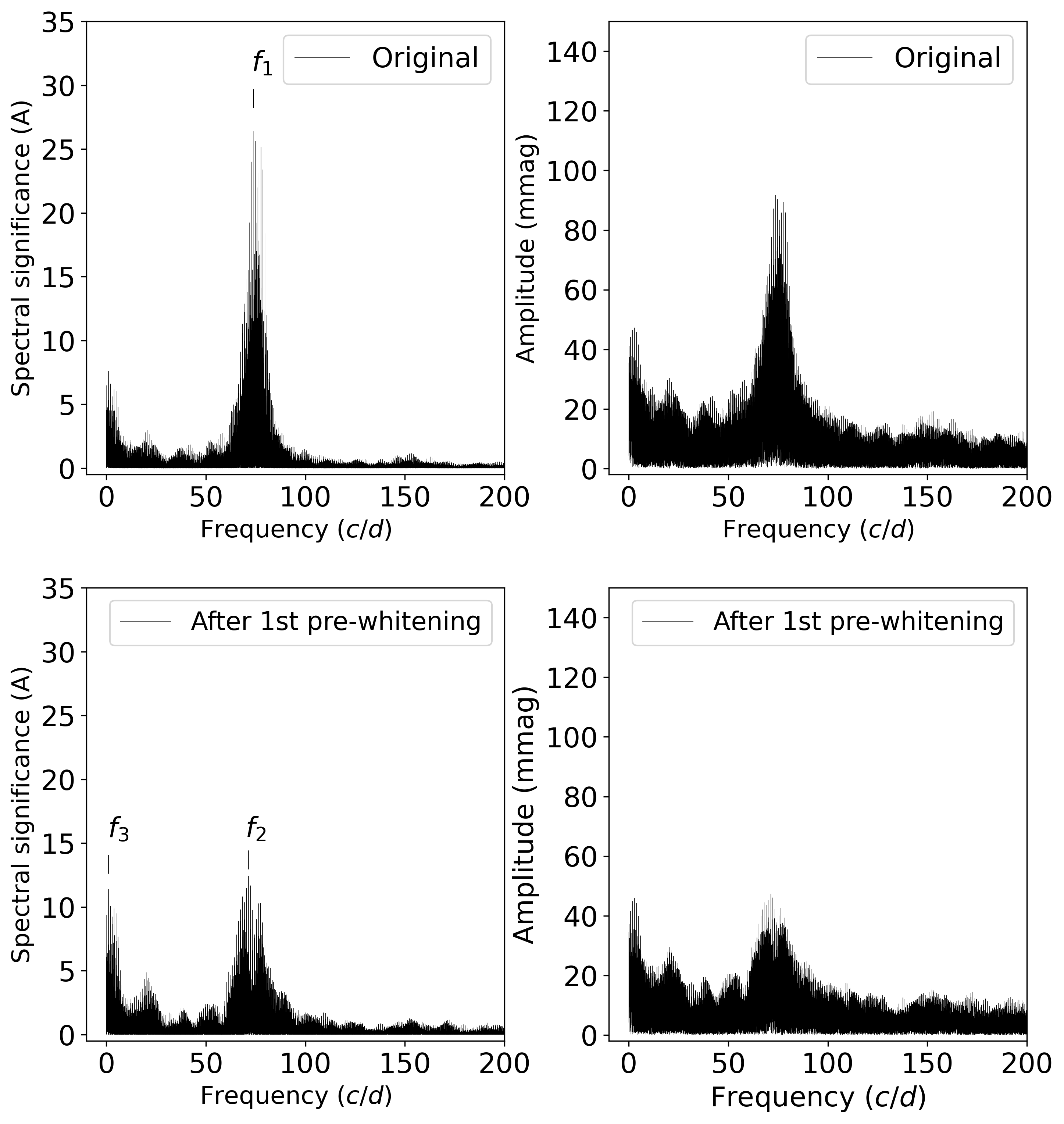}
\caption{Top panel: Phase-folded $u$-band light curve with the dominant period of $f_{1}$ = 73.7664 c/d (19.5211 min). We subtract the median value of $u$-band magnitude, 16.0562 mag, from the light curve. Middle panels: Significance spectrum (left) and DFT amplitude spectrum (right) of the high-cadence $u$-band photometry before pre-whitening. Bottom panels: Same as Middle panels but it is calculated after pre-whitening with the  frequency $f_{1}$. Non-negligible signal {\refbf (SNR $>$ 4)} at $f_{2}$ = 71.3855 c/d is found in the residual data and 1-cycle-per-day alias, $f_{3}$, is identified as well.}
\label{fig: Periodicity}
\end{figure}

\begin{figure}
\includegraphics[width=\linewidth]{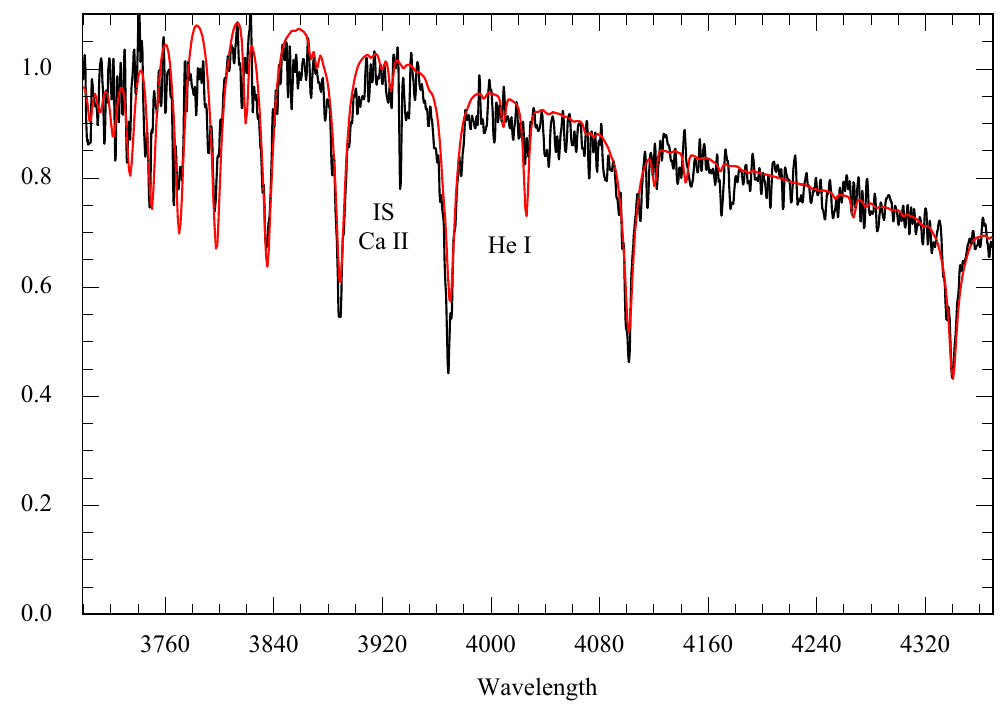}
\includegraphics[width=\linewidth]{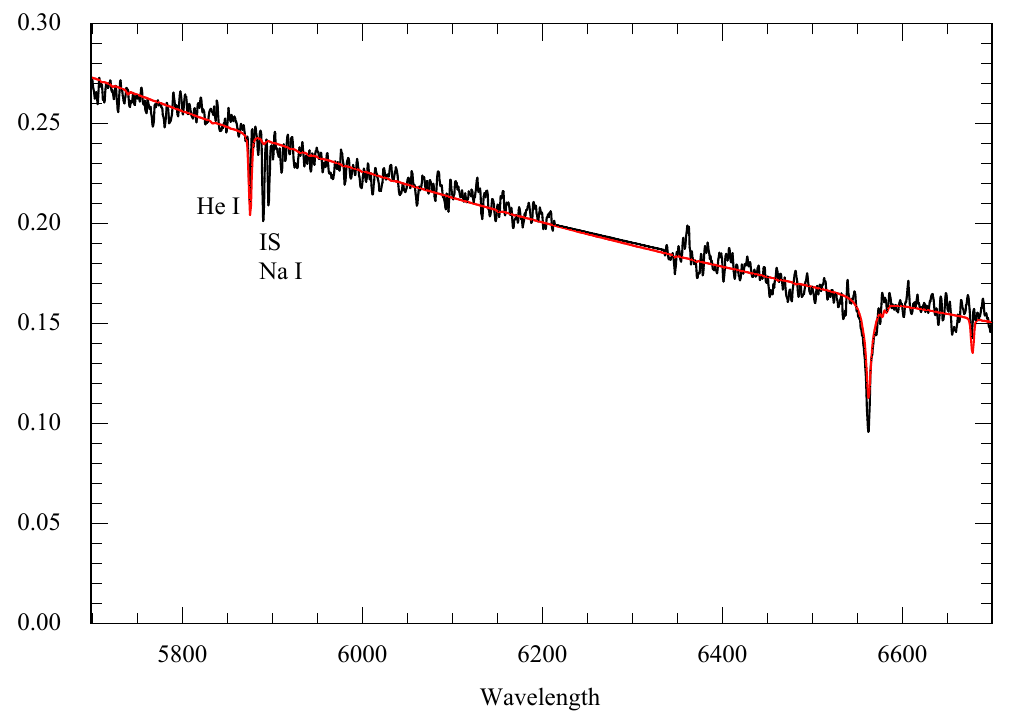}
\caption{The combined WiFeS blue (top) and red (bottom) spectrum of SMSS-BLAP-1 (black line) compared with the Munari synthetic spectrum (red line).} 
\label{fig: WiFeS spectrum and initial template matching}
\end{figure}

\begin{figure*}
\includegraphics[width=0.95\linewidth]{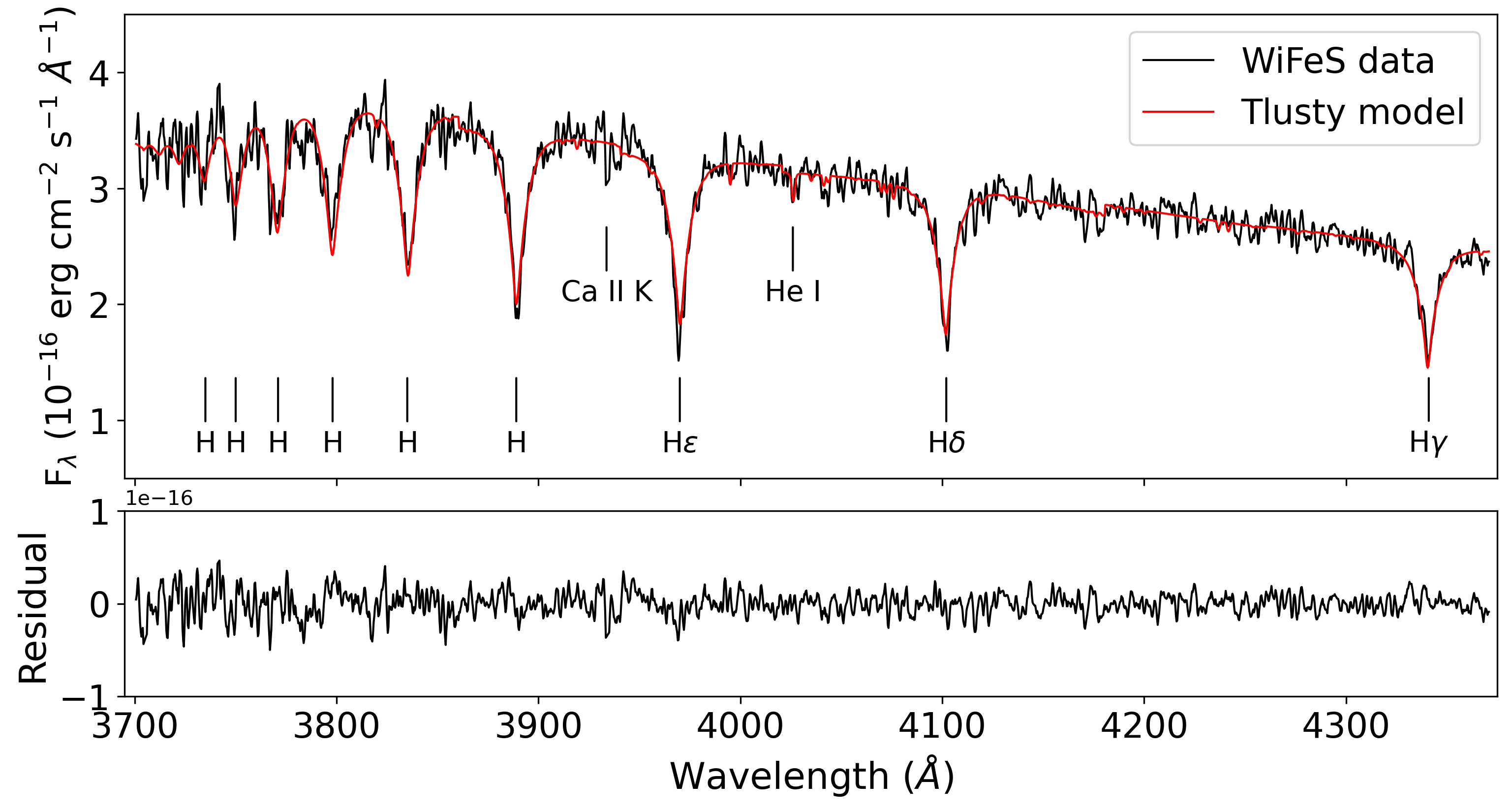}
\caption{Top panel: Comparison between the synthetic Tlusty/Synspec model spectrum (red line) and the co-added radial-velocity corrected B7000 WiFeS data (black line) for notable hydrogen and He I lines. {\refbf There is a hint of a non-stellar Ca II K line.} The best-fit model parameters are: $T_\mathrm{eff}$ = 29,000 K, $\log g$ (cm s$^{-2}$) = 4.66 and $\log$ $n$(He)/$n$(H) = -2.72 dex. Bottom panel: The residual spectrum between observation and model.} 
\label{fig: Stellar Parameter Verification with Tlusty and Synspec}
\end{figure*}

\subsection{Pulsation Period}
To identify a pulsation period and its amplitude, we perform a frequency analysis of the high-cadence lightcurve using the Period04 \citep{Lenz2005CoAst.146...53L} and the Sigspec \citep{Reegen2007A&A...467.1353R} software. The former is based on a classical Discrete Fourier Transform (DFT) for the detection of sinusoidal type signals. The latter computes the spectral significance levels (unbiased statistical estimator) for the DFT analytically through integration of the amplitude probability density function. To overcome the aliasing problem caused by non-equidistant sampling, it can also perform an anti-aliasing correction rather than a step-by-step pre-whitening technique. 

We find identical results with the two packages where the most dominant frequency is 73.7664$\pm$0.0005 cycles per day ($f_{1}$), which translates into 19.5211$\pm$0.0001 minutes. The uncertainty of the frequency is calculated from Monte Carlo simulation with Period04. The phase-folded $u$-band light curve in Figure \ref{fig: Periodicity} shows a sinusoidal oscillation (radial pulsation) with a peak-to-peak amplitude of {\refbf $\sim$0.2} mag, which is in the typical observed range of 0.1-–0.4 mag in optical bands (e.g., see Figure 1 of \citealt{Lin2022NatAs.tmp..217L}). More importantly, SMSS-BLAP-1 is located near the edge of the 8-to-20 min period gap, where we still have limited information in the current sparse BLAP samples. The existence of this discontinuity is questionable because it can be explained either as a natural consequence of formation channels and evolutionary pathways or as just an observational selection effect. \citet{Byrne2021MNRAS.507..621B} argue that the lack of observational discoveries in this region may be due to the mass distribution of the underlying pre-white dwarf population. However, the recent discovery of more BLAPs within the gap \citep{Lin2022NatAs.tmp..217L,Borowicz2023AcA....73....1B} suggests that they are more common than hitherto believed.

Next, we investigate observational evidence for other low-degree radial and/or non-radial pulsation modes in our BLAP,  as predicted by stellar evolution models of hot subdwarfs (e.g., \citealt{Romero2018MNRAS.477L..30R,2019A&ARv..27....7C}). We expect to detect non-radial pulsation in surface temperature in our $u$-band lightcurve, which is much sensitive to temperature variations than previous $I$- or $r$-band light curves. The significance and amplitude spectra before and after prewhitening of the $f_{1}$ signal are shown in the middle and bottom panels of Figure \ref{fig: Periodicity}, respectively. No additional significant peaks are present at or near integer multiples of the primary frequency. The remaining frequencies of $f_{2}$ (= 71.3855$\pm$0.001 c/d = 20.1722$\pm$0.0003 min) and $f_{3}$ (= 1.00445 c/d) are shown as vertical lines. Except for the 1-cycle-per-day alias, $f_{2}$ may be a non-radial mode with low amplitude ($\sim$0.09$\pm$0.006 mag) and sinusoidal shape in the light curve. However, it is unclear whether the second significant peak in the residual spectrum is genuine, as it may be due to flux contribution from a nearby non-variable source (see Section 2.2). Further observations and theoretical work are needed to definitively determine the nature of the pulsation modes for SMSS-BLAP-1.

\begin{figure}
\includegraphics[width=0.98\linewidth]{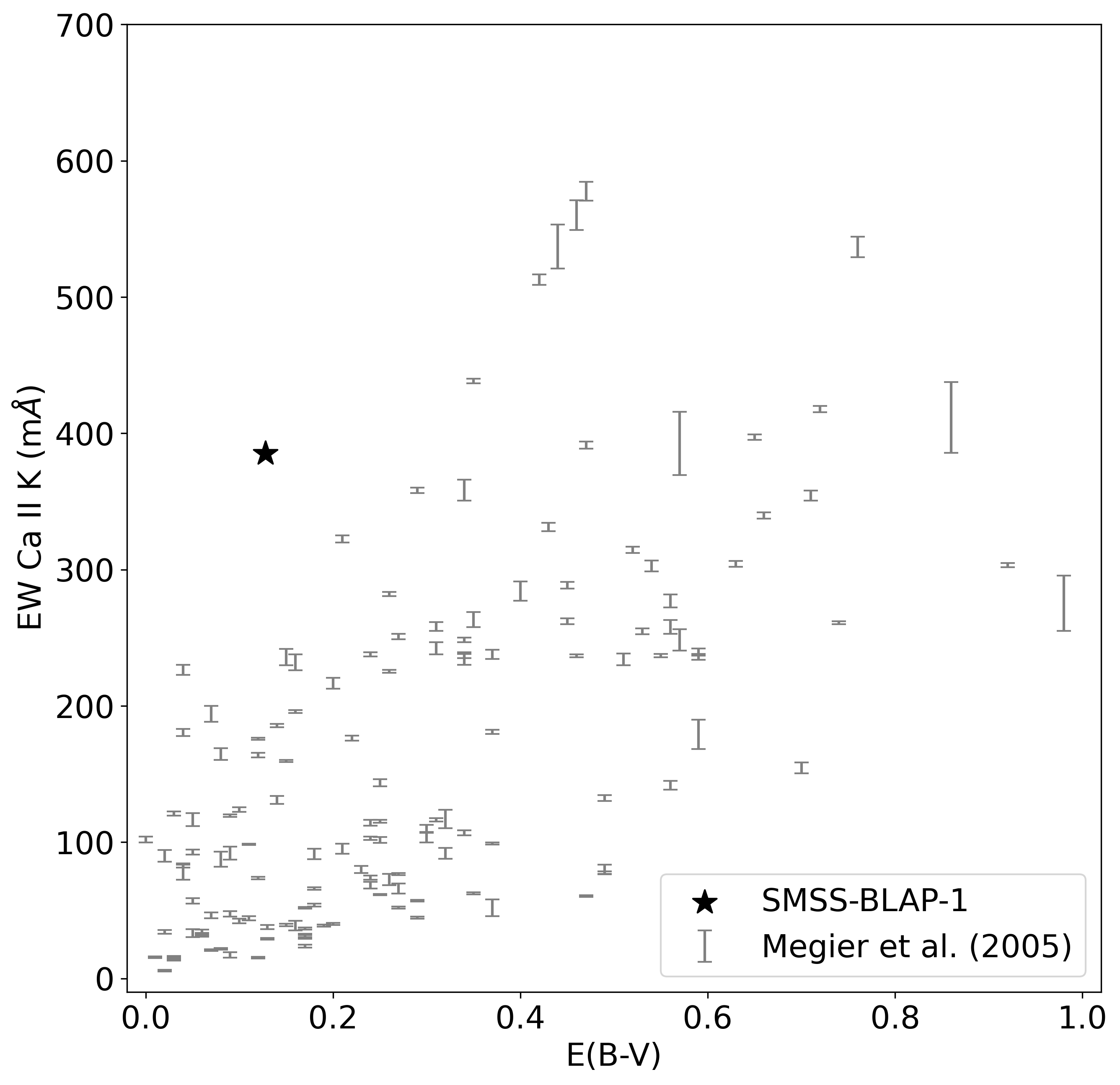}
\caption{Relation between Ca II K line strength and $E(B-V)$ for OB stars in our Galaxy \citep{Megier2005ApJ...634..451M} and SMSS-BLAP-1 (star).}
\label{fig: circumstellar material}
\end{figure}

\subsection{Stellar Parameters}
To derive the effective temperature and surface gravity, we initially compare the fluxed blue spectrum with a grid of Munari synthetic spectra \citep{Munari2005A&A...442.1127M} using the following criteria: 15,000 $<$ $T_\mathrm{eff}$ $<$ 50,000, 3.5 $<$ $\log g$ $<$ 5.5, and -1.0 $<$ [M/H] $<$ 0.0 dex. This comparison results in a good match with a template star possessing an effective temperature of 27,000 K, a surface gravity of $\log g$ = 4.4, a metallicity of [M/H] = -1.0, and $E(B-V)$ = 0.128. The measured radial velocity (RV) was found to be -55 km s$^{-1}$ with an uncertainty of up to 10 km s$^{-1}$. Cross-correlation of individual blue spectra against the mean spectrum indicates that the RV is constant across 90 minutes with an RMS uncertainty of 1.5 km s$^{-1}$. The signal-to-noise ratio of the individual spectra does not allow us to detect any changes in the line shapes of the hydrogen lines. At such a high temperature, the surface gravity and metallicity are not as well constrained as the temperature when fitting the blue spectrum. It is worth noting that the observed He I 4026 line is noticeably weaker in the star than in the model spectra, as is the He I 5876 line in the red spectrum (Figure \ref{fig: WiFeS spectrum and initial template matching}). This characteristic of weak He is a property of known subdwarf O and B stars.

To obtain precise atmospheric parameters, we conduct spectral analysis using the online platform XTgrid \citet{Nemeth2019ASPC..519..117N}. XTgrid utilises the non-local thermodynamic equilibrium model atmosphere structure and synthetic spectra computed with Tlusty/Synspec \citep{Hubeny1995ApJ...439..875H,Hubeny2017arXiv170601859H}, covering the relevant parameter space for hot subdwarf stars. We restrict the models to include only the chemical elements H, He, C, N, and O. The iterative procedure minimising $\chi^{2}$, simultaneously optimises all free parameters to derive the best-fit model. For the blue-only spectrum, the adopted model parameters are as follows: {\refbf $T_\mathrm{eff}$ = 29,020$^{+193}_{-34}$ K, $\log g$ = 4.661$^{+0.008}_{-0.143}$ (cm s$^{-2}$), $\log$ $n$(He)/$n$(H) = -2.722$^{+0.057}_{-0.074}$ and RV = -53.1$^{+1.7}_{-1.7}$ km s$^{-1}$(Figure \ref{fig: Stellar Parameter Verification with Tlusty and Synspec}). The uncertainties in these parameters are determined from 68\% confidence intervals}. Notably, our BLAP exhibits significant helium deficiency compared to those reported by OGLE (-0.41$\sim$0.55 dex; \citealt{Pietrukowicz2017NatAs...1E.166P}) and ZTF (-2.0$\sim$-2.4 dex; \citealt{Kupfer2019ApJ...878L..35K}). One plausible explanation could be the gravitational settling of helium (e.g., \citealt{Hu2010A&A...511A..87H}). Moreover, the SED fit suggests a relatively modest reddening value with E$(B-V)$=0.126$\sim$0.128.

A prominent feature in the spectra are strong, narrow, interstellar lines of Ca II K (3933.7\AA) and Na I D (5889.95/5895.924\AA). The rest-frame equivalent width (EW) of the Ca II K line and Na I D is approximately 390 and 400 m\AA, respectively. The strength of the Na line is slightly more uncertain than the Ca II measurement due to telluric Na emission. Observations of Galactic field O and B stars by \citet{Megier2005ApJ...634..451M} suggest that the measured strength of the interstellar lines in this object would correspond to an $E(B-V)$ between 0.3 and 0.7 (see Figure \ref{fig: circumstellar material}), significantly exceeding the reddening estimates from \citet{Schlegel1998ApJ...500..525S}. Given this evidence, it is {\refbf possible} that most of the line EWs originate from circumstellar material (CSM) rather than ISM. The detection of significant CSM around BLAPs may constrain their recent evolutionary history.

\begin{figure}
\includegraphics[width=\linewidth]{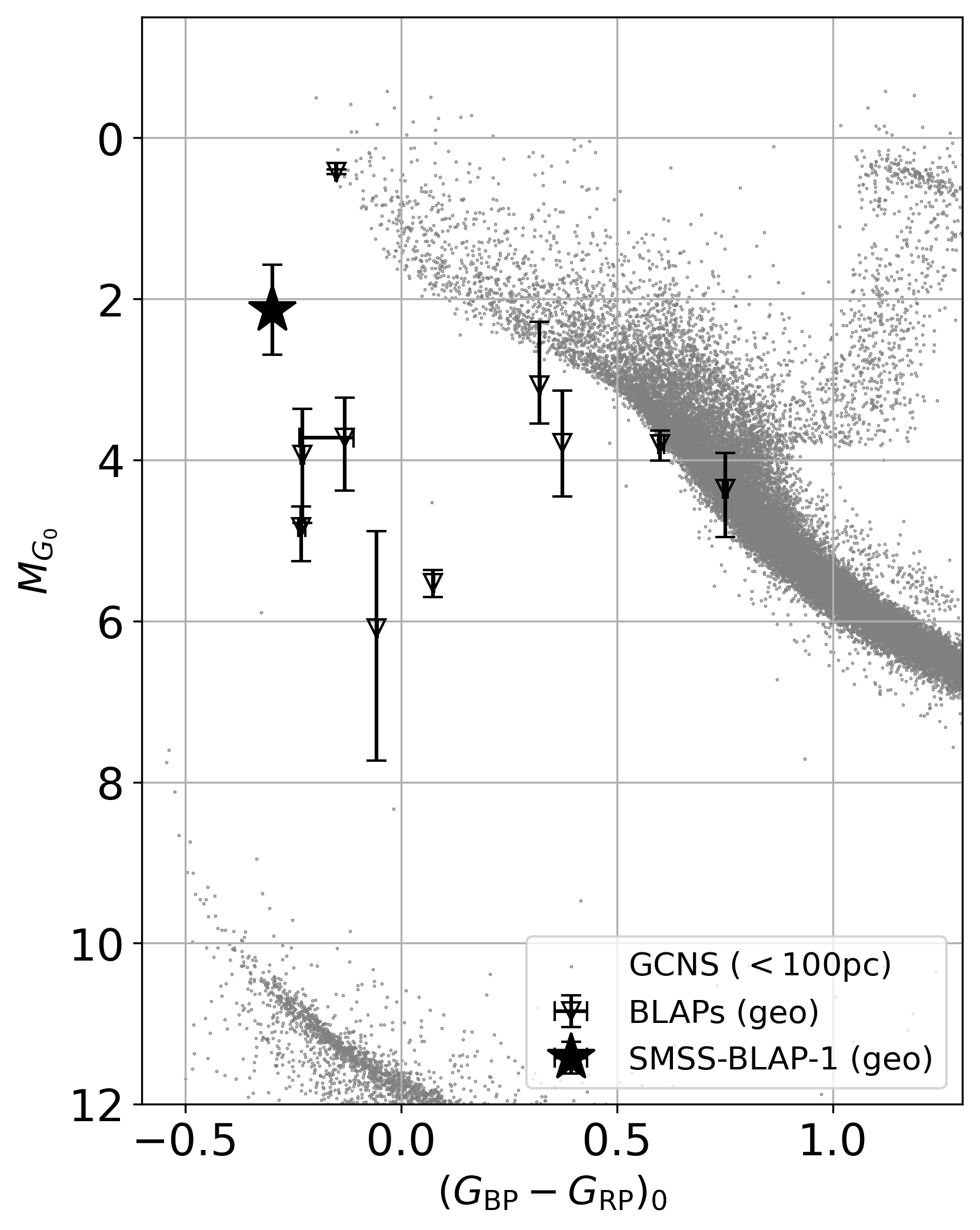}
\caption{Un-reddened Gaia DR3 colour-magnitude diagram for known 10 BLAPs (open inverse triangle), SMSS-BLAP-1 (star), and nearby stars within 100 pc (grey). Error bars represent the 16th to 84th percentile confidence interval.}
\label{fig: BLAP CaMD}
\end{figure}

\section{GAIA DR3 View of BLAPs}\label{sec:discussion}
\subsection{Hertzsprung-Russell diagram of BLAPs}
To understand characteristic properties of BLAPs in common, we describe the Gaia Data Release 3 \citep{GAIADR32016A&A...595A...1G, GAIA2022arXiv220800211G} view of their current evolutionary path in the colour-absolute magnitude diagram (CaMD). We cross-matched the compiled list of known BLAPs to Gaia sources with full astrometric solutions using a matching radius of 1.0\arcsec. As recommended by the Gaia Collaboration \citep{Lindegren2021A&A...649A...4L,Riello2021A&A...649A...3R, Fabricius2021A&A...649A...5F}, we then apply a set of quality cuts on the astrometric (\texttt{ruwe} < 1.4 \& \texttt{parallax}/\texttt{parallax\textunderscore error} > 1 \& \texttt{ipd\textunderscore frac\textunderscore multi\textunderscore peak} $\leq$ 2) and photometric measurements (\texttt{phot\textunderscore g\textunderscore mean\textunderscore mag} is not NULL \& \texttt{phot\textunderscore bp\textunderscore mean\textunderscore mag} is not NULL \& \texttt{phot\textunderscore rp\textunderscore mean\textunderscore mag} is not NULL \& \texttt{phot\textunderscore bp\textunderscore rp\textunderscore excess\textunderscore factor} < 1.3). Next, we retain clean objects with any distance estimates derived from two posterior probability distributions, $d_\mathrm{geo}$ and $d_\mathrm{photogeo}$, under the assumption that the source is a single star in our Galaxy (see \citealt{Bailer-Jones2021AJ....161..147B} for details). Table \ref{tab:tab1} includes unique names, the Gaia DR3 ID, Gaia three-band photometry, extinction $A_{G}$, reddening $E(G_{BP}-G_{RP})$, parallax $\overline{\omega}$ and its signal-to-noise ratio $\overline{\omega}_\mathrm{SNR}$, and distances for the final eleven BLAPs. The parallax measurement for our BLAP is 0.1239$\pm$0.0569 mas with SNR = 2.2, which is only a weak constraint, but combined with priors from Galactic structure implies a distance of roughly 6--10 kpc. Although the parallax SNR of nearly all BLAPs are low (SNR$<$5), we primarily use the geometric rather than the photogeometric distance to avoid dependence on colour and magnitude priors.

The dust de-reddened \(G\) absolute magnitude and \(G_\mathrm{BP}-G_\mathrm{RP}\) colour of each star were obtained as 
\begin{eqnarray*}
M_{G_{0}} &=& G - A_{G} - 5 \log_{10} d_\mathrm{geo} + 5, \\
(G_{BP} - G_{RP})_{0} &=& G_{BP} - G_{RP} - E(BP-RP),
\end{eqnarray*} where $A_{G}$ and $E(BP-RP)$ are the extinction and reddening estimates inferred by GSP-Phot Aeneas \citep{Creevey2022arXiv220605864C}, respectively. Figure \ref{fig: BLAP CaMD} shows the comparison between our clean sample of 11 BLAPs and the Gaia Catalogue of Nearby Stars within 100~pc (\citealt[GCNS;][]{GCNS2021A&A...649A...6G}). Despite having large uncertainties in distance estimations, the known BLAPs cover a very wide range of colour and luminosity in the CaMD, implying the difficulty of searching for new BLAPs wihtout light curves. We find that SMSS-BLAP-1 is the most luminous and bluest BLAP so far except for the binary star HD133729. 

\begin{figure}
\includegraphics[width=0.97\linewidth]{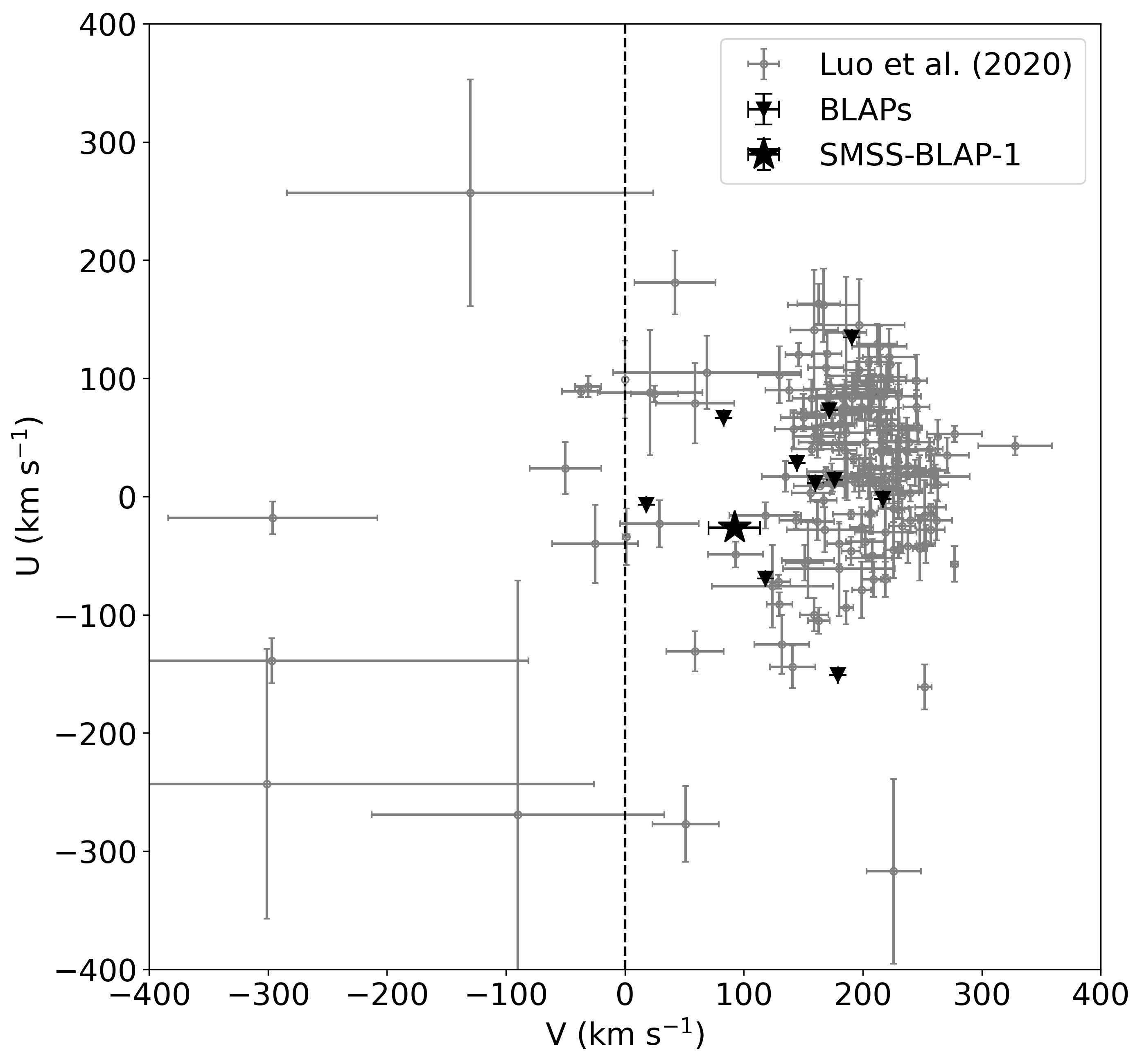}
\includegraphics[width=0.98\linewidth]{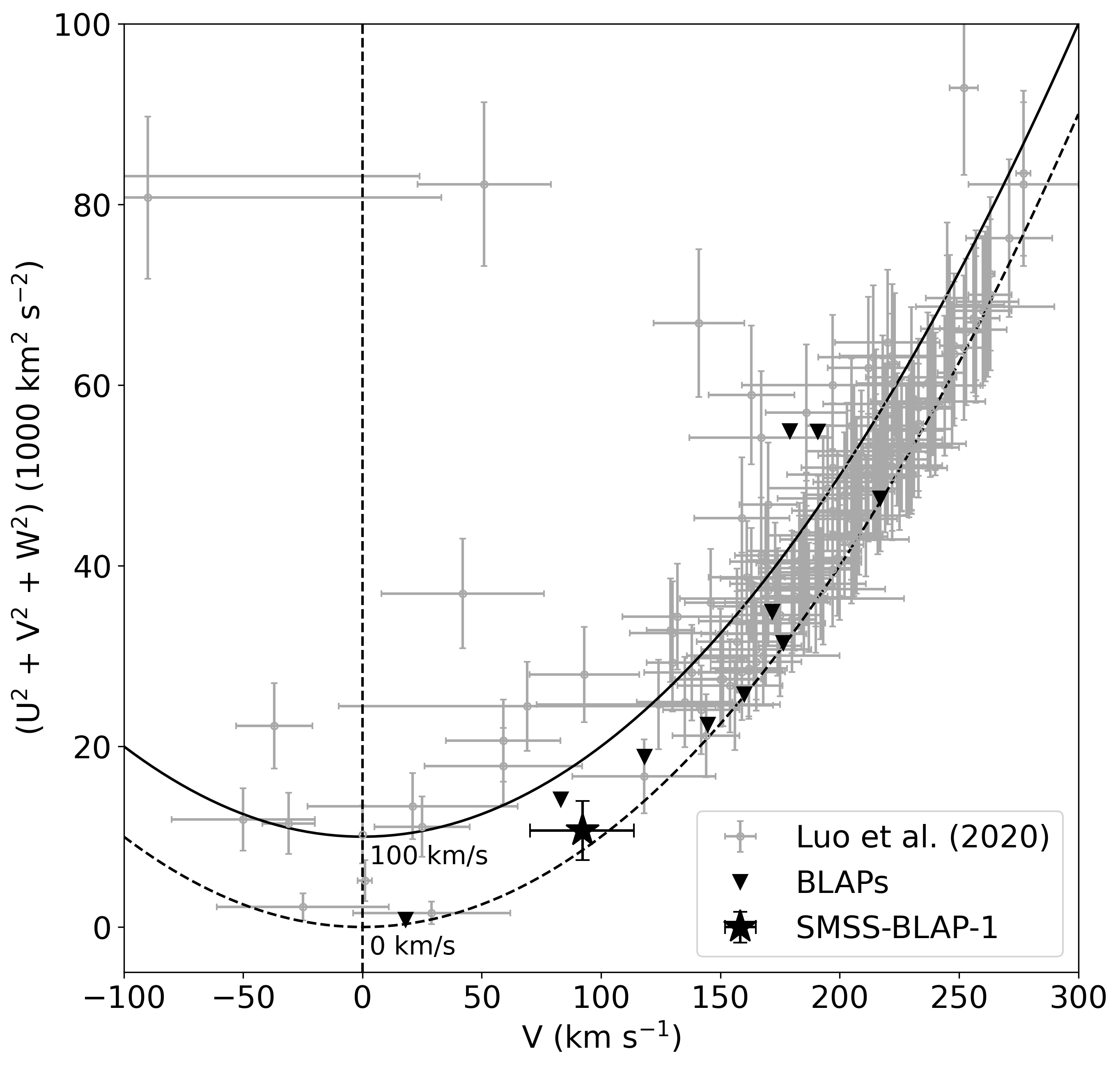}
\caption{Top panel: $U$-$V$ velocity diagram for 182 hot subdwarfs from \citet{Luo2020ApJ...898...64L} and BLAPs. Bottom panel: Galactic rotational velocity $V$ against the total kinetic energy 2$E_\mathrm{kin}/m = U^2 +V^2 + W^2$ for the same sample. The two curves show the velocities perpendicular to Galactic rotation at $v_{\perp}$ = $(U^{2} + W^{2})^{1/2}$ = 0 km s$^{-1}$ and 100 km s$^{-1}$, respectively.}
\label{fig: BLAP V vs. Ekin}
\end{figure}

\subsection{Kinematic properties of BLAPs}
We further use Gaia DR3 data to infer the Galactic positions and velocities along with kinematic properties for BLAPs listed in Table \ref{tab:tab1}. From the position, parallax and proper motion, we compute the 3D Cartesian position ($X, Y, Z$) and velocity ($U, V, W$) components in the Galactocentric frame with the astropy python package \citep{2013A&A...558A..33A}, where a right-handed system is used. In the right-handed system, ($U$, $V$, $W$) are positive in the direction toward the Galactic Centre, Galactic rotation and North Galactic Pole. Following the recent guidelines by \citet{2023A&A...674A..39G}, we assume that the Galactocentric radius of the Sun is set at 8.277 kpc \citep{2022A&A...657L..12G} and its local circular velocity is 219 km s$^{-1}$ \citep{Bovy2015ApJS..216...29B}. We also adopt the updated measurement of the Sun's height above the Galactic midplane \citep{Bennett2019MNRAS.482.1417B} and its peculiar motion \citep{2010MNRAS.403.1829S}: $Z_{\odot}$ = $20.8\pm0.3$ pc and ($U_{\odot}, V_{\odot}, W_{\odot}$) = (11.1$^{+0.69}_{-0.75}$, 12.24$^{+0.47}_{-0.47}$, 7.25$^{+0.37}_{-0.36}$) km s$^{-1}$, respectively. Table \ref{tab:tab2} presents a summary of the positions and velocities of 11 BLAPs, including cases where radial velocity is not considered. 

SMSS-BLAP-1 is further from the Galactic plane ($Z=-1.463$~kpc) than other known BLAPs, but it is closer to the Galactic centre. The uncertainty in the distance estimate is a primary factor affecting the 6D vector. Considering the 16th and 84th percentiles of the distance posterior, we see small variations on the order of $\pm$0.1--0.4 kpc in the $Y$ and $Z$ positions and $\pm$4--5 km s$^{-1}$ in the $U$ and $W$ velocities. Significant differences are only seen in the $X$ position and $V$ velocity component, but the results are not affected. The derived value of tangential ($v_\mathrm{tan}$) and total ($v_\mathrm{tot}$) velocities suggest that our BLAP is kinematically a thick disc star, as those typically have 40 $\leq$ $v_\mathrm{tan}$ $\leq$ 180 km s$^{-1}$ and 70 $\leq$ $v_\mathrm{tot}$ $\leq$ 180 km s$^{-1}$ (e.g., \citealt{Bensby2014A&A...562A..71B,2023A&A...674A..39G}). We also compare the kinematic properties of hot subdwarfs \citep{Luo2020ApJ...898...64L} and BLAPs. As shown in Figure \ref{fig: BLAP V vs. Ekin}, our BLAP has a relatively slow rotational velocity and exhibits prograde motion ($V > 0$). The low value of its total kinetic energy, 2$E_\mathrm{kin}/m$ = $U^{2} + V^{2} + W^{2}$, indicates that our BLAP is in a circular orbit. Lastly, we see that it is a kinematically cool star as $v_{\perp}$ = $(U^{2} + W^{2})^{1/2}$ is very close to the iso-velocity curve of 0 km s$^{-1}$.

\begin{table*}
\caption{Gaia DR3 photometry and astrometry of 11 BLAPs with clean flags.}
\begin{tabular}{ccccccccccc}
\hline \noalign{\smallskip}
Name & Source ID & $G$ & $G_\mathrm{BP}$  & $G_\mathrm{RP}$ & $A_\mathrm{G}$ & $E(BP-RP)$ & $\overline{\omega}$ & $\overline{\omega}_\mathrm{SNR}$& $d_\mathrm{geo}$ & $d_\mathrm{photogeo}$ \\ 
     & (Gaia DR3) & (mag) & (mag) & (mag) & (mag) & (mag) & (mas) &  & (pc) & (pc)  \\
\noalign{\smallskip} \hline \noalign{\smallskip}
OGLE-BLAP-1 & 5254042907771535616 & 17.479 & 17.615 & 17.240 & 0.0012 & 0.0007 & 0.139 & 1.96 & 5462$^{+1956}_{-1420}$ & 9253$^{+1787}_{-1394}$ \\ 
OGLE-BLAP-9 & 4063563189835683968 & 15.555 & 15.773 & 15.168 & 0.0097 & 0.0053 & 0.4142 & 10.98 & 2228$^{+219}_{-169}$ & 2528$^{+179}_{-203}$ \\ 
OGLE-BLAP-10 & 4042004343740886272 & 17.151 & 17.308 & 16.986 & 0.0021 & 0.0011 & 0.2078 & 2.16 & 6529$^{+1597}_{-1989}$ & 8264$^{+1042}_{-1075}$ \\ 
OGLE-BLAP-14 & 4048980156591219456 & 16.754 & 16.760 & 16.713 & 0.5184 & 0.2782 & 0.5444 & 6.53 & 1905$^{+408}_{-212}$ & 2266$^{+686}_{-417}$ \\ 
ZTF-BLAP-1 & 3032063709154939392 & 16.513 & 16.536 & 16.461 & 0.0013 & 0.0007 & 0.6091 & 10.53 & 1571$^{+129}_{-116}$ & 1674$^{+136}_{-137}$ \\ 
ZTF-BLAP-2 & 4073369076865715456 & 18.806 & 18.864 & 18.919 & 0.0045 & 0.0024 & 0.6298 & 2.61 & 3486$^{+3936}_{-1484}$ & 980$^{+204}_{-136}$ \\ 
ZTF-BLAP-3 & 4198057990399279360 & 17.581  & 17.608 & 17.620 & 0.4052 & 0.2171 & 0.2191 & 2.26 & 4454$^{+2127}_{-1022}$ & 4170$^{+1243}_{-946}$ \\ 
ZTF-BLAP-4 & 4484346014846313472 & 17.249 & 17.243 & 17.218 & 0.2926 & 0.1559 & 0.211 & 2.97 & 4439$^{+1584}_{-901}$ & 4256$^{+1433}_{-827}$ \\ 
HD133729 & 6210817001175183104 & 9.111 & 9.112 & 9.068 & 0.3613 & 0.1935 & 2.1357 & 73.36 & 463$^{+6}_{-6}$ & 462$^{+6}_{-6}$ \\ 
TMTS-BLAP-1 & 473013789224384896 & 16.907 & 17.202 & 16.424 & 0.0482 & 0.0266 & 0.3208 & 4.97 & 3160$^{+996}_{-588}$ & 4370$^{+691}_{-614}$ \\ 
\noalign{\smallskip} \hline \noalign{\smallskip}
SMSS-BLAP-1 & 6737383019655229696 & 16.476 & 16.389 & 16.637 & 0.092 & 0.0495 & 0.1239 & 2.18 & 7063$^{+2058}_{-1622}$ & 8925$^{+1157}_{-1135}$ \\ 
\noalign{\smallskip} \hline
\end{tabular}
\label{tab:tab1}
\end{table*}

\begin{table*}
\caption{Space positions and Galactic velocities of BLAPs in Galactocentric frame.}
\begin{tabular}{cccccccc} 
\hline \noalign{\smallskip}
Name & $X$ & $Y$ & $Z$ & $U$ & $V$ & $W$ & $v_\mathrm{tan}$ \\ 
     & (kpc) & (kpc) & (kpc) & (km s$^{-1}$) & (km s$^{-1}$) & (km s$^{-1}$) & (km s$^{-1}$) \\ 
\noalign{\smallskip} \hline \noalign{\smallskip}
OGLE-BLAP-1 & -6.6 & -5.2 & -0.2 & -151.0 & 179.1 & -7.8 & 225.5 \\ 
OGLE-BLAP-9 & -6.1 & 0.1 & -0.1 & 14.0 & 176.4 & 12.0 & 59.7 \\ 
OGLE-BLAP-10 & -1.8 & -0.5 & -0.6 & -6.9 & 18.1 & -21.0 & 159.0\\ 
OGLE-BLAP-14 & -6.4 & 0.0 & -0.2 & 11.3 & 160.1 & -1.0 & 69.1 \\ 
ZTF-BLAP-1 & -9.3 & -1.2 & -0.0 & -69.4 & 118.2 & 8.3 & 15.3 \\ 
ZTF-BLAP-2 & -4.9 & 0.5 & -0.6 & 28.4 & 144.7 & 25.5 & 41.0 \\
ZTF-BLAP-3 & -4.3 & 1.8 & -0.8 & 66.5 & 83.1 & -53.1 & 173.4 \\
ZTF-BLAP-4 & -5.0 & 2.9 & 0.8 & 134.6 & 190.8 & -17.9 & 126.0 \\ 
HD133729 & -7.9 & -0.2 & 0.2 & -2.1 & 216.9 & 20.5 & 23.8 \\ 
TMTS-BLAP-1 & -10.8 & 1.8 & 0.2 & 73.2 & 171.8 & 5.7 & 12.8 \\
\noalign{\smallskip} \hline \noalign{\smallskip}
SMSS-BLAP-1 & -1.4$^{+2.0}_{-1.6}$ & 0.6$^{+0.2}_{-0.1}$ & -1.5$^{+0.3}_{-0.4}$ & -24.5$^{+4.7}_{-3.7}$ & 92.3$^{+30.9}_{-39.2}$ & 38.3$^{+5.8}_{-4.6}$ & 156.3 \\
\noalign{\smallskip} \hline
\end{tabular}
\label{tab:tab2}
\end{table*}

\section{Summary}\label{sec:summary}
We present the newly discovered BLAP (SMSS J184506-300804) from the SMSS DR2 data. The identification of {\refbf one more star} in this category is still important, as BLAPs are a very rare class of pulsating variables. This discovery was facilitated by the short cadence of $uv$ band exposure pairs in the Main Survey colour sequence component of the SkyMapper Southern Survey \citep{Onken2019PASA...36...33O}.

A high-cadence follow-up monitoring revealed a period of 19.5211 minutes in the $u$ band, observed for the first time in $u$ band. The new object also indicates that the number of systems in the 8–20 minute period gap could be increased. In our combined spectrum of 9$\times$600 seconds, we estimate the effective temperature, surface gravity, and helium abundance of the BLAP to be {\refbf $T_\mathrm{eff}$ = 29,020$^{+193}_{-34}$ K, $\log g$ = 4.661$^{+0.008}_{-0.143}$ (cm s$^{-2}$), and $\log$ n(He)/n(H) = -2.722$^{+0.057}_{-0.074}$ dex (He-poor)}. These values are sufficient to confirm the nature of the BLAP as a low-gravity BLAP, but they are not precise because the best-fit parameters vary depending on the pulsation phase. Additionally, the SNR of the individual spectra is insufficient to resolve changes in the line shapes of the hydrogen and He I lines.

Our BLAP exhibits two notable features: (i) an unusual periodic signal in the residual lightcurve, which could be indicative of a non-radial pulsation mode, and (ii) the presence of excess absorption in Ca II K and Na I D lines suggesting the presence of CSM. The origin of the periodic signal is still unclear, but we speculate whether the neighbouring star may be involved. The existence of CSM around BLAPs is significant because it raises the question of its origin. Given that most BLAPs are believed to be the result of helium core burning, one possible explanation is that the CSM could be from the envelope lost due to the impact of the ejecta from SN Ia (e.g., \citealt{Meng2020ApJ...903..100M}). The CSM may remain around the progenitor system for a very long time, depending on how it loses its angular momentum. One important question is whether our BLAP is the only one with an abnormally strong CSM, or whether other published BLAP spectra also exhibit similar levels of interstellar absorption lines; unfortunately, the literature spectra do not cover the relevant spectral regime to readily answer this question. The relatively high-latitude location of our BLAP with its low reddening value gives us an advantage in identifying CSM signatures via excess atomic absorption beyond the level expected for the ISM from the maximum line-of-sight reddening. 

In the fourth data release of SkyMapper (SMSS DR4; \citealt{Onken2024arXiv240202015O}), we will extend our search for Southern BLAPs. DR4 will not only more than triple the number of available image SMSS images, but extend in particular the sky coverage in $uv$ observations as part of the SMSS Main Survey colour sequence. These enhancements will enable a systematic search for low- and large-amplitude blue variables.

\section*{Acknowledgements}
The authors are grateful to Zhanwen Han and Xiangcun Meng for helpful discussions. SWC acknowledges support from the National Research Foundation of Korea (NRF) grants, No. 2020R1A2C3011091 and No. 2021M3F7A1084525 funded by the Ministry of Science and ICT (MSIT). This research was also supported by Basic Science Research Program through the NRF funded by the Ministry of Education (RS-2023-00245013). CAO was supported by Australian Research Council (ARC) Discovery Project DP190100252. The national facility capability for SkyMapper has been funded through ARC LIEF grant LE130100104 from the Australian Research Council, awarded to the University of Sydney, the Australian National University, Swinburne University of Technology, the University of Queensland, the University of Western Australia, the University of Melbourne, Curtin University of Technology, Monash University and the Australian Astronomical Observatory. SkyMapper is owned and operated by The Australian National University's Research School of Astronomy and Astrophysics. The survey data were processed and provided by the SkyMapper Team at ANU. The SkyMapper node of the All-Sky Virtual Observatory (ASVO) is hosted at the National Computational Infrastructure (NCI). Development and support the SkyMapper node of the ASVO has been funded in part by Astronomy Australia Limited (AAL) and the Australian Government through the Commonwealth's Education Investment Fund (EIF) and National Collaborative Research Infrastructure Strategy (NCRIS), particularly the National eResearch Collaboration Tools and Resources (NeCTAR) and the Australian National Data Service Projects (ANDS). This work has made use of data from the European Space Agency (ESA) mission {\it Gaia} (\url{https://www.cosmos.esa.int/gaia}), processed by the {\it Gaia} Data Processing and Analysis Consortium (DPAC, \url{https://www.cosmos.esa.int/web/gaia/dpac/consortium}). Funding for the DPAC has been provided by national institutions, in particular the institutions participating in the {\it Gaia} Multilateral Agreement. We thank Fabiola Marino for executing some of the spectroscopic observations. This research has also used the services of \url{www.Astroserver.org} under reference Z9Y8AT.


\section*{Data availability}
The data underlying this article were accessed from the SkyMapper website  \url{http://skymapper.anu.edu.au}. The derived data generated in this research will be shared on reasonable request to the corresponding author.



\bibliographystyle{mnras}
\bibliography{SMSS_DR2_BLAP}







\bsp	
\label{lastpage}
\end{document}